\documentclass[sigconf, nonacm]{acmart}

\usepackage{xpatch}

\makeatletter
\xpatchcmd{\ps@firstpagestyle}{Manuscript submitted to ACM}{}{\typeout{First patch succeeded}}{\typeout{first patch failed}}
\xpatchcmd{\ps@standardpagestyle}{Manuscript submitted to ACM}{}{\typeout{Second patch succeeded}}{\typeout{Second patch failed}}    \@ACM@manuscriptfalse
\makeatother

\setcopyright{none}

\AtBeginDocument{%
  \providecommand\BibTeX{{%
    \normalfont B\kern-0.5em{\scshape i\kern-0.25em b}\kern-0.8em\TeX}}}

\usepackage{enumitem}
\usepackage{amsthm,amsmath,amssymb}
\usepackage[ruled,vlined]{algorithm2e}
\usepackage{mathrsfs}
\usepackage{multirow}
\usepackage{lipsum}
\usepackage{subcaption}
\usepackage{multicol}
\usepackage{lipsum}
\usepackage{pifont}
\usepackage{url}
\usepackage{hyperref}

\SetCommentSty{mycommfont}
\SetKwInput{KwInput}{Input} 
\SetKwInput{KwOutput}{Output}

\begin{document}
\title{LSM-OPD: Boosting Scans in LSM-Trees by Enabling Direct  Computing on Compressed Data}


\author{Jianfeng Huang}
\affiliation{%
  \institution{Harbin Institute of Technology}
  \city{Harbin}
  \state{China}}
\email{jfhuang.research@gmail.com}

\author{Ziyao Wang}
\affiliation{%
  \institution{Harbin Institute of Technology}
  \city{Harbin}
  \state{China}}
\email{zywang.research@gmail.com}

\author{Lin Yuan}
\affiliation{%
  \institution{Harbin Institute of Technology}
  \city{Harbin}
  \state{China}}
\email{lyuan.research@gmail.com}

\author{Jiajie Wen}
\affiliation{%
  \institution{Harbin Institute of Technology}
  \city{Harbin}
  \state{China}}
\email{jjwen.research@gmail.com}

\author{Yihao Cao}
\affiliation{%
  \institution{Harbin Institute of Technology}
  \city{Harbin}
  \state{China}}
\email{yhcao.research@gmail.com}

\author{Dongjing Miao}
\affiliation{%
  \institution{Harbin Institute of Technology}
  \city{Hangzhou}
  \state{China}}
\email{miaodongjing@hit.edu.cn}

\author{Yong Wang}
\affiliation{%
  \institution{Gauss Laboratory, Huawei Company}
  \city{Hangzhou}
  \state{China}}
\email{wangyong308@huawei.com}

\author{Jiahao Zhang}
\affiliation{%
  \institution{Gauss Laboratory, Huawei Company}
  \city{Hangzhou}
  \state{China}}
\email{cupermanrose@gmail.com}


\begin{abstract}
Scan-based operations, such as backstage compaction and value filtering, have emerged as the main bottleneck for LSM-Trees in supporting contemporary data-intensive applications. For slower external storage devices, such as HDD and SATA SSD, the scan performance is primarily limited by the I/O bandwidth (\textit{i.e.,} I/O bound) due to the substantial read/write amplifications in LSM-Trees. Recent adoption of high-performance storage devices, such as NVMe SSD, has transformed the main limitation to be compute-bound, emerging the impact of computational resource consumption caused by inefficient compactions and filtering. However, when the value size increases, the bottleneck for scan performance in fast devices gradually shifts towards the I/O bandwidth as well, and the overall throughput across all types of devices undergo a dramatic reduction.

This paper addresses the core issues by proposing LSM-OPD, a \underline{L}og-\underline{S}tructured \underline{M}erge-\underline{O}rder-\underline{P}reserving \underline{D}ictionary encoding scheme that enables direct computing on compressed data within LSM-Trees.
It first enables key-value-separated data flushing to disk in a densely encoded columnar layout, ideally with few bytes for a large string value (\textit{e.g.,} $1024$  bytes), thereby significantly alleviating the frequent I/O requests caused by intensive scans. Then, it is capable of offloading the costly scan-based operations on large values, including compaction and value filtering, 
to lightweight dictionaries due to the order-preserving property. And SIMD-based vectorization can now be employed to maximize the evaluating performance on modern multi-core processors, further breaking the compute-bound limitations in LSM-trees. Extensive experiments demonstrate the superior efficiency of LSM-OPD in processing various workloads  that involve intensive scan-based operations on diverse modern storage devices.

\end{abstract}

\maketitle
\renewcommand{\thefootnote}{}



\section{Introduction}

The Log-Structured Merge (LSM)-Tree is widely adopted as the underlying storage engine by numerous database systems including RocksDB \cite{RocksDB}, LevelDB \cite{LevelDB}, Cassandra \cite{Cassandra}
and HBase \cite{Hbase}, serving for various data-intensive applications to enhance write performance with a tolerable increase in lookup latency. However, recent data-intensive applications, such as the Internet of Things (IoT) and 5G, real-time fraud detection and prompt decision-making, which involve complex data analysis concurrent with intensive data insertions (\textit{i.e., }hybrid transactional and analytical processing), pose an urgent demand for high-performance processing of scan-based predicates in LSM-Trees~\cite{dayan2018dostoevsky,sarkar2020lethe,saxena2023real,choi2020ourrocks}.
On the other hand, LSM-Trees heavily rely on frequent backstage scan-based compaction operations to persist data modifications and reclaim garbage during the data propagation to deeper levels on disk, which itself is resource-intensive but essential for maintaining the lookup efficiency. 
To make matters worse, the backstage compactions may compete fiercely for computational resources with front query processing, especially when other scan-based predicates are under evaluating, thereby undermining the overall throughput of LSM-trees. As a result, scan-based operations have emerged as the main bottleneck for LSM-Trees in supporting modern data-intensive applications.

\begin{figure}[h]
\centering
\includegraphics[width=1\linewidth]{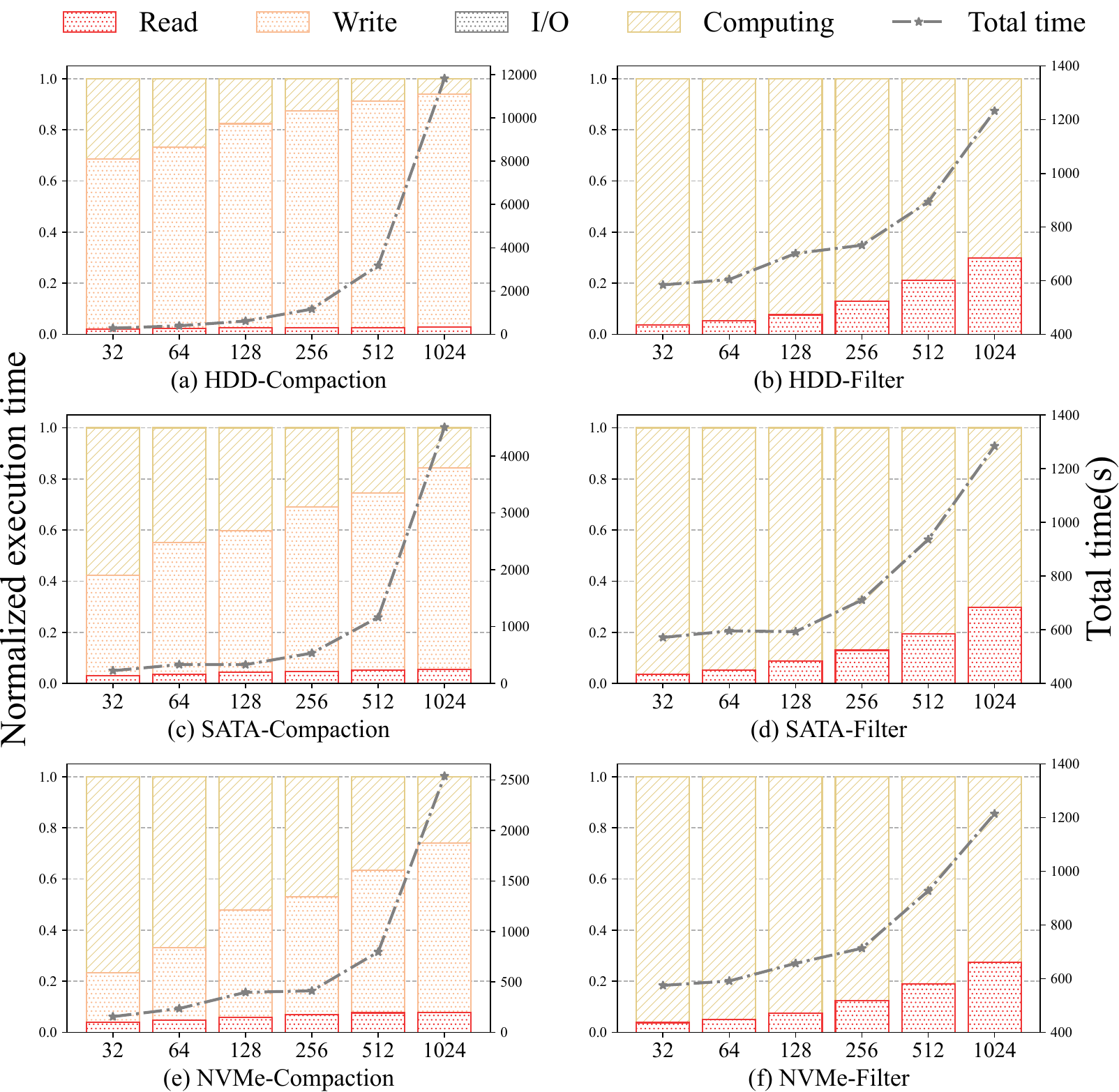}
\caption{The time breakdown analysis of LSM-based compaction and filter operation for different value size in RocksDB~\cite{RocksDB} on different storage devices (Upon $6.4\times10^7$ key-value pairs insertions and performed 10 filters).}
\label{bd}

\end{figure}

However, the scan-based operations are heavily limited by the inherent read/write amplifications within the design space of LSM-Trees. Typically, in modern LSM engines (\textit{e.g.,} LevelDB~\cite{LevelDB}, Rocks-DB~\cite{RocksDB}), a compaction operation comprises of seven stages: \textit{file retrieval, reading, decoding, merging, filtering, encoding,} and \textit{writing}, while a value filtering operation involves the first five stages, which are both resource-intensive.
On one hand, due to the overlapping key ranges across different levels, scan-based operations require intensive I/Os to access (extra writing cost for compactions) multiple SSTs (Sorted String Tables) across multiple levels that contain the target key ranges.
On the other hand, massive CPU resources are required to decode and reorder the entries from multiple SST files, thereby filtering out unqualified values and discard stale versions. As shown in Figure~\ref{bd}.(a)(c)(e), different types of storage devices encounter distinct bottlenecks in processing scan-based compactions based on their I/O performance when the value size is relatively small, \textit{i.e.,} I/O bound for HDD and SATA SSD, and compute-bound for NVMe SSD. As for filter processing, Figure~\ref{bd}.(b)(d)(f) have emerged the negative effects
of inefficient filter computations.
However, Figure~\ref{bd} also shows that, as the size of the value (generally large strings) increases, the bottlenecks for scan-based operations gradually shift towards I/O
bandwidth and dramatically increase overall latency in all types of storage devices (see the dotted histograms and lines in Figure~\ref{bd}), regardless of the hardware performance. This is mainly because more frequent I/Os are required to process the scan-based operations within the same key range in LSM-Trees, and the overhead of string comparison will dramatically increase as the size of the string becomes larger.

\noindent\underline{\textit{\textbf{Challenges.}}} A straightforward solution to reduce the value size and alleviate massive I/O requests caused by intensive scans is leveraging compression techniques that are widely adopted in modern columnar database systems. 
However, directly applying common compression techniques can't thoroughly address the core issues for LSM-Trees. Because more decoding computation is traded for less I/Os during the processing of scan-based operations, which will further raise the compute-bound for fast storage devices and undermine the overall performance. More importantly, most compression techniques lack support for direct modifications on encoded data, resulting in high recoding computational cost during the compactions in LSM-Trees. 
Therefore, designing an inherently embedded compression scheme that can universally boost scan-based operations for diverse performance bounds in LSM-Trees still presents significant challenges, as it requires a thorough optimization on underlying storage layout, query evaluation scheme, and compaction policy within the design space of LSM-Trees.  

To address these challenges, we outline the following
desiderata for an ideal compression scheme embedded within LSM-Trees: \textbf{\normalsize{\Large\textcircled{\scriptsize{$D1$}}}}~Suitable for key-value pairs compression, especially for large strings, and achieves a relatively high compression ratio, \textbf{\normalsize{\Large\textcircled{\scriptsize{$D2$}}}}~Supports lightweight recoding and avoids introducing additional complexity during the compactions within LSM-Trees, \textbf{\normalsize{\Large\textcircled{\scriptsize{$D3$}}}}~Enables direct filtering on compressed data, thus eliminating additional computational resource consumption of decompression. 

\noindent\underline{\textit{\textbf{Motivations.}}} In this paper, we revisit the order-preserving dictionary (OPD) encoding scheme and find that it provides promising potentials to meet all the requirements for radically boosting scan-based operations in LSM-Trees due to the following observations:
\textbf{\normalsize{\Large\textcircled{\scriptsize{$O1$}}}}~It establishes a bijective order-preserving mapping from an arbitrary infinite source domain (large
string value domain) to a fixed-size encoded domain (often small integers, and can be further compressed using bit-packed encoding),  providing high compression ratio for strings when the NDVs (number of distinct values) is relatively low; \textbf{\normalsize{\Large\textcircled{\scriptsize{$O2$}}}}~It has the potential to  offload expensive compactions from large values to lightweight dictionaries and enable direct merge on encoded domain, thereby eliminating the expensive recoding costs during the compactions within LSM-Trees; \textbf{\normalsize{\Large\textcircled{\scriptsize{$O3$}}}}~It enables direct filtering on the encoded domain due to the order-preserving property, and further facilitates SIMD-based vectorization 
to maximize the evaluating performance on modern
multi-core processors.  

However, the order-preserving property typically comes at a substantial cost. Especially when facing intensive data changes, where the upcoming value domain is unknown and hard to predict. Because the dense encoded domain can be violated when there is no more ``code space” for new values, a more complex code domain must be pre-allocated to maintain the OP property, thus losing support for dense SIMD-based vectorization; otherwise, the entire code domain have to be reconstructed, resulting in substantial compute costs and temporary stalls.  
However, we observe that, when embedded within the design space of LSM-Trees, an OPD 
can be constructed for each \textit{memtable} during data flushes to disk. In this way,
OPD reconstructions can be only executed over foregone fixed value domains due to adherence to the \textit{out-of-place} paradigm, and the costs are naturally concealed by the backstage compaction operations
and amortized
at the file grain.

\noindent\underline{\textit{\textbf{Contributions.}}} These observations provide strong motivations
for us to seamlessly embed the OPD-based encoding scheme within the design space of LSM-Trees, and leads to the following contributions in this paper: 

\begin{itemize}[leftmargin=*]
\item{We make the first attempt to thoroughly boost scan-based operations for diverse performance bounds within LSM-Trees by enabling direct computing on compressed data.}

\item{We first significantly mitigate the I/O-bound within LSM-Trees by introducing a novel log-structured merge-order-preserving dictionary encoding scheme with a densely compressed columnar key-value-separated storage layout.}

\item{We further break the compute-bound within LSM-Trees by offloading the scan-based operations to lightweight OPDs, which enables more efficient compactions with reduced complexity and fast vectorized filter processing directly on encoded data.}

\item{We integrate everything and build LSM-OPD, which achieves enhanced overall throughput in various workloads involving intensive scan-based operations.
}

\end{itemize}

\section{Backgrounds}
\label{bg}

\begin{figure}[h]
\centering
\includegraphics[width=1\linewidth]{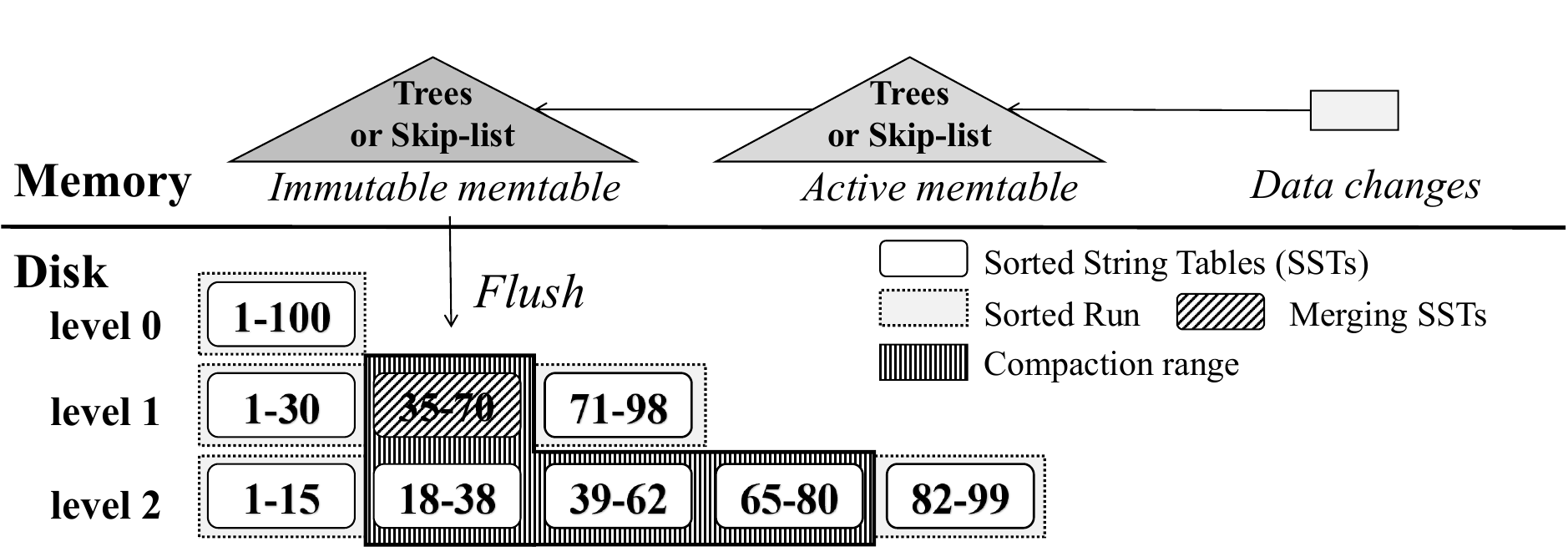}
\caption{Partitioned \textit{Leveling} compaction policy}
\label{MP}

\end{figure}

\noindent\textbf{LSM-Tree Basics and Compaction.}
The LSM-tree (Log-structured Merge-tree) has been widely adopted in the storage layer of modern NoSQL systems (\textit{e.g.,} key-value storage RocksDB~\cite{RocksDB}).  Different from traditional index structures that apply in-place updates, LSM-Trees defer data file writes and buffer data modifications in a memory-resident segment, while employing a tree structure or
\textit{skip-list} to maintain the order of inserted records. These modifications are then propagated to the immutable disk files through sequential and batched I/Os. As shown in Figure~\ref{MP}, the immutable files on disk are organized into \textit{sorted runs} in different levels. The entries in each \textit{sorted run} are sorted based on the index keys, which can be stored in a single file or alternatively partitioned into several smaller files known as \textit{Sorted String Tables (SSTs)}.
With the continuous writing of data changes, the files on disk will accumulate over time. And the query performance tends to degrade, as it may require accessing more files with overlapping key range in order to locate target entry with a given key. 

To alleviate this problem, the disk segment is organized into $L$ logical levels of increasing size with a size ratio of $T$.  And the files are merged during the data propagation to deeper levels by a backstage process called
\textit{compaction}. In practice,
the \textit{leveling} compaction policy is commonly adopted due to its superior read performance, despite the increased computational cost during compactions. As shown in Figure~\ref{MP}, it strictly limits a single sorted run for each level. In \textit{leveling} merge,
an SST at level $L$ will be merged multiple times with incoming SSTs at level $L-1$ until it fills up. Then it will
be merged with multiple SSTs at level $L+1$ that share the overlapping
key ranges with it. Where each merge operation involves the aforementioned seven costly stages, resulting in intensive resources consumptions. As a result, the \textit{leveling} merge suffers higher write amplifications, but provides better read performance than the \textit{tiering} merge, as it necessitates multiple merges of a full SST with several SSTs in the subsequent level, but ensuring non-overlapping key ranges among the SSTs within each level after merges. 

\noindent\textbf{Operations in LSM-Trees.} LSM-Tree based systems follow the \textit{out-of-place} data ingestion paradigm, where the data modifications are first buffered in a memory-resident structure and then be propagated to deeper levels on disk  through sequential and batched disk I/Os. The \textit{out-of-place} data ingestion paradigm and the multi-level layout lead to a distinct implementation of classic database operations in LSM-Trees. 

\begin{itemize}[leftmargin=*]
\item $insert (key, value)$: Inserting a key-value pair in LSM-based engines typically only requires locating it in the ordered memory-resident segment based on the sorting structure (\textit{e.g., skip-list}). The simplified data insertion enables efficient transactional processing,
as it does not incur any random and fragment disk I/Os.  

\item $delete(key)$: Deletion is typically considered a secondary citizen in LSM-Trees~\cite{sarkar2020lethe}. When a key-value pair requires to be deleted, a tombstone with the same key will be inserted to the memory-resident structure. Physical deletion will be triggered when the tombstone encounters its owner during data propagation to deeper levels by the merge operations, resulting in a considerable delete persistence latency. 

\item $update(key, value)$: Updates are typically executed by directly inserting a new version of the KV pair. And the stale version will be reclaimed during compactions, since only the entry with
the most recent timestamp can survive.

\item $point\_lookup(key)$: Since entries are updated in an \textit{out-of-place} manner, multiple versions of an entry with the same key may exist across multiple levels. A point lookup searches for the appropriate version of the entry by traversing from the smallest level to the largest. This process is usually accelerated by maintaining a bloom filter for each file in memory.
It terminates upon encountering the first entry with a matching key.
\item $range\_lookup(key_{low}, key_{high})$: A range lookup has to find the most recent versions of all entries within the target key range. It is typically implemented based on a scan iterator, which will simultaneously examine across multiple levels, identifying and discarding stale versions with the same key.

\item $filtering(Value_{\{conditions\}})$: Evaluating a filter in LSM-Trees typically requires scanning all SST files across all levels to distinguish the key-value pairs whose values satisfy the filtering conditions, as well as discarding the stale versions, which is highly resource-intensive. 

\end{itemize}

In modern data-intensive applications (\textit{e.g.,} HTAP workloads), LSM-Tree based engines may be required to process concurrent queries that include a mix of writes and reads, with recent data accessed by OLTP-style queries ($inserts$, $deletes$, $updates$ and $point$  $\_lookups$, $short\ range\_lookups$), and both recent and older data accessed by OLAP-style queries ($long
range\_lookups$, and $filtering$). This poses an urgent demand for boosting scan-based operations in LSM-based systems, as a large number of scan-based front queries and backstage compactions may compete fiercely for compute resources at the same time, leading to a dramatically reduced overall throughput.

\noindent \textbf{Order-preserving Dictionary Encoding.}
Order-preserving dictionary encoding is a lossless compression scheme that maps the arbitrary infinite source domain to a compact fixed-size encoded domain (commonly in small integer representations) while strictly preserving the original order of the symbols. Formally, given an ordered sequence of input values: $S = \{S_1,S_2,...,Sn\}$ with a total ordering relation $s_i \prec s_j$ for $i<j$, an OPD constructs a bijective function $\mathcal{E} : S \leftrightarrow\{0,1,...,m-1\} $ such that:
$
\forall \  s_{i}, s_{j} \in S, \quad s_{i} \prec s_{j} \iff \mathcal{E}(s_{i}) < \mathcal{E}(s_{j})
$
, where $m \le n$ is the number of distinct symbols in $S$. The encoded integers maintain the original symbol ordering, enabling value comparison and range operations directly on compressed data. And it enables cascading compression by assigning minimal $log_2m$ bits (\textit{i.e.,} in a bit-packed manner) to each symbol in the encoded domain, further facilitating SIMD-based vectorization.

\begin{figure*}[h]
\centering
\includegraphics[width=1\linewidth]{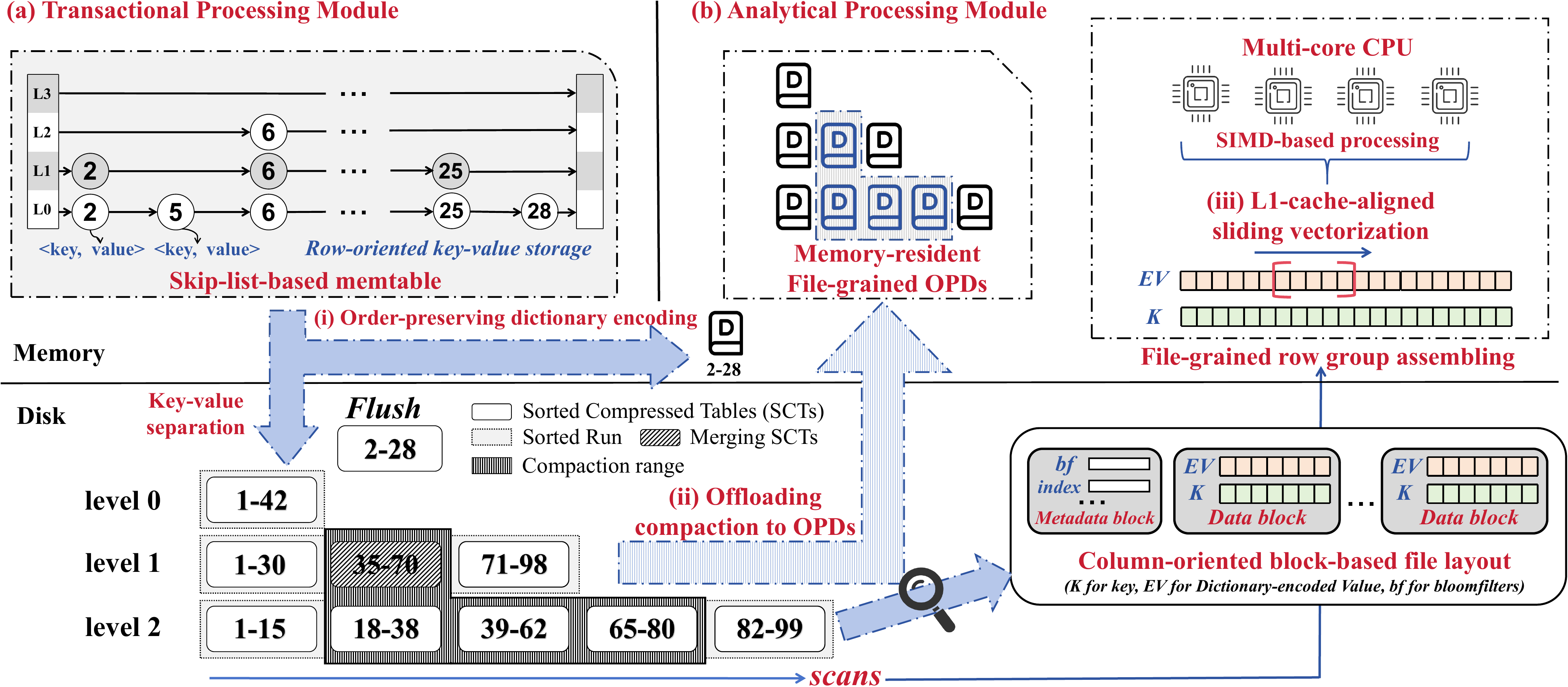}

\caption{Overview of \textit{Log-Structured Merge Order-preserving Dictionary (LSM-OPD)}}
\label{OV}
\end{figure*}

\section{System Designs of LSM-OPD}
\noindent\textbf{System Overview.}
As shown in Figure~\ref{OV}, LSM-OPD comprises of an in-memory buffering component for fast transactional processing and an on-disk persisting component for efficient scan-based analytics.
Specifically,
there are two main memory-resident modules within LSM-OPD, one for transactional processing and the other for analytical processing. Data insertions are first buffered in a row-oriented lock-free \textit{skip-list}-based \textit{memtable} to  accelerate transactional processing throughput (Figure~\ref{OV}~(a)). Upon reaching its maximum capacity, a \textit{memtable} will be frozen and compressed into an immutable \textit{Sorted Compressed Table} (\textit{SCT}) that is ready to be flushed to disk using the OPD encoding. During this process, the layout will be transformed into a key-value-separated column-oriented manner (Figure~\ref{OV} (i)). 
While the file-grained OPDs will be maintained in the memory-resident analytical processing module to further boost subsequent scan-based operations, since they are typically small in practice. 

As for the disk-based component, the space is organized into multiple logical levels by the compaction
process as well. Once a level reaches its maximum capacity, a \textit{SCT} in this level will be merged with multiple \textit{SCTs} at the next level following the  \textit{leveling} compaction policy. And the heavy computation will be offloaded to lightweight OPDs, leading to a minor compaction overhead (Figure~\ref{OV} (ii)). As for scan-based predicate evaluations, the block-wise key-value columns in SCTs will be assembled into file-grained row groups in the memory-resident analytical processing module without decompression, facilitating further SIMD-based vectorized processing directly on compressed data (Figure~\ref{OV} (iii)).

Overall, LSM-OPD seamlessly embeds the order-preserving dictionary encoding scheme within the design space of LSM-Trees, universally boosting the scan-based operations, including compactions and filters in LSM-Trees, for diverse performance bounds by enabling direct computing on compressed data.

\noindent\textbf{Memory-resident buffering component.}
Data insertions are first buffered in the memory-resident \textit{skip-list}-based \textit{memtables}, following the \textit{out-of-place} data ingestion paradigm. The key-value pairs are stored at the bottom of the \textit{skip-list} in a row-oriented form to facilitate the transactional processing. This has a negligible impact on analytic performance, as the \textit{memtables} are significantly smaller compared to the on-disk persisting component.
Upon reaching its maximum capability, a \textit{memtable} will be frozen and stop receiving new changes any more, indicating that the source domain of the \textit{memtable} is now fixed. And the construction of an order-preserving dictionary is now transformed to a simple and lightweight sorting problem, where 
the large string values will be replaced by its position index in the original value domain based on alphabetic order (\textit{i.e.,} encoded value \textit{ev} ), leading to a densely encoded \textit{SCT} that is prepared to be flushed to disk.
In this way, the heavy cost of constructing an OPD for the entire dynamic value domain are amortized
at the \textit{SCT} grain and
concealed by the backstage flush operation.
This process will also transform the row-oriented layout to the column-oriented key-value-separated layout in \textit{SCTs}, further facilitating vectorized analytical processing.

\noindent\textbf{On-disk persisting component.} As shown in Figure~\ref{OV}, the \textit{SCT} files are grouped into multiple logical levels by the \textit{leveling} compaction policy as well, where each level contains a single sorted run. 
In each \textit{SCT}, keys and encoded values are organized into small column chunks in blocks (4 kb in practice). And the file metadata, such as block-wise bloom filters, key ranges and offsets, are stored in extra blocks. The block-based management facilitates \textit{point\_lookup} and \textit{short\_range lookup} by pruning unnecessary block retrievals, while have negligible impact on analytical performance since all blocks are still consecutively stored, thereby ensuring purely sequential scans as well. When a \textit{SCT} is full, it will trigger a compaction with the \textit{SCTs} in the next level that share the overlapping key ranges with it. This computation will be offloaded to memory-resident lightweight OPDs, and generate new dense dictionaries for \textit{SCTs} after compactions~(\S\ref{OP}).

\section{Operations in LSM-OPD}

\subsection{Basic Transactional Operations}
\noindent\textbf{Data Modifications.} Transactions are processed in LSM-OPD basically following the classic \textit{out-of-place} data ingestion paradigm described above in the background (\S\ref{bg}).
In the memory-resident buffering component, the stale versions will not be physically deleted immediately. Instead, their lifetime intervals, denoted as [\textit{Creation\_time} ($T^{C}$), \textit{Deletion\_time} ($T^{D}$)], will be recorded to ensure compliance with the \textit{snap-shot} isolation level and prevent unauthorized accesses.
If the stale version has already been flushed to disk, a tombstone indexed by the same key will be inserted into the \textit{skip-list} and be propagated to deeper levels on disk by the backstage compactions to indicate its deletion.
The delayed update mechanism prevents the frequent reconstructions of OPDs, and delegate them
to the backstage compactions in batches, thereby amortizing the heavy computations at the file grain.
Since all files in the \textit{on-disk} persisting component are immutable and only the compaction process can change the file layout, expensive fine-grained lifetime information is no longer needed. Instead, an accessible file snapshot, which includes the physical addresses of each currently visible \textit{SCT} and the corresponding OPD, is assigned to each read transaction according to its timestamp. This lightweight MVCC (multi-version concurrent control) scheme also ensures the accurate retrieval of key-value versions consistent with the \textit{snapshot} isolation level as well.

\noindent\textbf{Lookup operations.} We focus on transactional   \textit{point\_lookup(key)} and  \textit{short range \_lookup($key_{low}$, $key_{high}$)} first.
During the search for specific key-value pairs, the \textit{memtables} are accessed with preference. 
This process involves $O(logM)$ entry lookup to the newest legal version based on the \textit{skip-list} index, where $M$ represents the total number of KV pairs stored in the \textit{memtable}.
If there is no appropriate version that satisfies the query condition in the  \textit{memtable}, the \textit{on-disk} persisting component is then accessed for further search. 
Specifically, the search on disk will initiate at level 0 and gradually proceed to deeper levels, accessing \textit{SCTs} containing the target key. Then, it will first check the bloom filters in the metadata blocks to prune the unnecessary data block accesses.
Since all values are encoded, an $O(1)$ lookup within the corresponding memory-resident OPD is required to decode the original values after block retrievals. 
For a point lookup,
if the search process encounters the tombstone of this KV pair first, which indicates that this KV pair has already been marked as deleted, the search will return an empty result.
Otherwise, the search process will terminate as soon as it finds the appropriate version of the KV pair in the \textit{memtable}, or encounters the latest version of it in a \textit{SCT}, since there is only one correct version living at the time the read query is processed.

As for a range lookup, we employ a similar iterator-based scan implementation like  RocksDB, where the search process will check across all levels at the same time
to discard the stale versions and ensure that no qualified KV pairs within the target key range is omitted.
And for a \textit{long range\_ lookup} that may require to access multiple \textit{SCTs} within the same level, it will trigger a bulk read to directly load the entire \textit{SCT} file into memory to facilitate subsequent process. As a result, the LSM-OPD basically remains the same high efficiency of point lookup and range lookup indexed by keys as the original LSM-Trees without any compression. Since the OPD-encoded value also refers to the offset of the original value in the memory-resident dictionaries and will not introduce extra heavy decoding costs to lookup operations. And the  denser block layout may enable more efficient system cache for data blocks, which may benefit the intensive compaction and lookup operations ~\cite{sarkar2021constructing}.

\newcommand{\tabincell}[2]{\begin{tabular}{@{}#1@{}}#2\end{tabular}} 
\begin{table}[h]
\setlength{\tabcolsep}{3pt}
\renewcommand{\arraystretch}{1}
\caption{Summary of used terms in this paper}
\label{term}
  \begin{center}
 \resizebox{\linewidth}{!}{
    \begin{tabular}
    {|c|l|c|c|}
    \hline
Term & Definition & Units & Reference value \\
\hline
$N$ & number of total inserted key-value pairs& 
entries & $2^{24}$\\
\hline
$l_i$ & level of $file_i$ & levels & $\le5$ \\
\hline
$F$ & prefixed size of each file & MB & $32$\\
\hline
$T$ & size ratio between adjacent levels & times & $\le10$ \\ 
\hline
$D_i$ & number of distinct value in $SCT_i$& entries & $10^{5}$\\
\hline
$S_K$ & size of keys& bytes & $16$\\
\hline
$S_V$ & size of uncompressed values& bytes & $64$\\
\hline
$S_O$ & size of OPD-encoded values& bytes & $4$\\
\hline
$C_K$ & average merge sort cost of keys & IPB & $1$\\
\hline
$C_C$ & average copy cost of unit-size data& IPB & 0.3\\
\hline
$C_E$ & average heavy compress cost of unit-size value& IPB & 50\\
\hline
$C_D$ & average heavy decompress cost of unit-size value& IPB & 20\\
\hline
$C_S$ & average string comparison cost of unit-size value& IPB & 1\\
\hline
$r$ & selectivity of a filter& percentages & 1\\
\hline
$S_I$ & size of processing data for a SIMD instruction& bytes & 512\\
\hline
    \end{tabular}
    }
    \label{terms}
  \end{center}
\end{table}

\subsection{Boosting Scan-based Operations}
\label{OP}

\subsubsection{\textbf{OPD-based \textit{Leveling} Compaction.}}

The core idea of LSM-OPD's design is to boost scans in LSM-Tree by enabling direct computation on compressed data. Compaction is one of the most essential scan-based operations within the design space of LSM-Tree, which greatly influences the overall performance by determining its update cost, delete persistence latency, lookups latency, space amplification, and write amplification~\cite{sarkar2021constructing,  dayan2018dostoevsky}. Compaction operations are resource-intensive and may compete fiercely for CPU resources with front query processing under intensive HTAP workloads. Thus, a more lightweight compaction scheme is urgently demanded by LSM-based systems. LSM-OPD achieves this by enabling direct compactions on compressed data, offloading the expensive computation from large string values to more lightweight OPDs. 

Specifically, the OPD-based \textit{leveling} compaction 
process is shown in Algorithm \ref{alg:algorithm1}. First, the key columns and value columns from $n$ merging \textit{SCTs} are assembled into  key-value pairs with file ID labels in memory. Then, a standard merge sort based on key order is executed to identify and reclaim stale versions, resulting in a long merged sequence of $<k',ev,s_i>$ tuples. The long sequence is then divided into $n'$ subsequences based on the prefixed \textit{SCT} file size. Until now, all values remain encoded according to the previous dictionaries within each subsequence; however, the order-preserving property is compromised as a result of the altered value domain. Thus, a merge of OPDs is required to rebuild a bijective order-preserving mapping in each merged subsequence. To achieve that, we construct a reverse index based on the original dictionaries. The reverse index maps an original value $v$ in $V_j$ to a set of its encoded values along with the corresponding \textit{SCT} ID, since the same value may be stored across multiple \textit{SCTs}. A \textit{RBTree} (\textit{std::map}) is employed to order the distinct values within the reverse index and thus construct the updated OPDs $O'$, where the duplicate string comparisons are pruned based on the reverse index in advance. 

As a result, the new OPDs $O'$ of merged \textit{$SCT's$} are rebuilt at the reduced cost of $O(\sum_{i=1}^n D_ilogD_i)$, where $D_i$ represents the number of distinct values (\textit{NDV}) in $SCT_i$. Then, an index table is built for each $SCT'_j$ based on the reverse index and the new OPD $O'_j$, which maps a tuple $<ev, s_i>$ to an updated encoded value $ev'$. And now each tuple $<k',ev,s_i>$ in the merged $SCT_j'$ can be deterministically mapped to the new encoded key-value pair $<k',ev'>$ at a constant $O(1)$ cost based on the index table, thereby efficiently completing the merge operation. As shown in Algorithm \ref{alg:algorithm1}, the updated encoding results are directly stored into key-value separated columns instead of being reassembled into key-value pairs, preparing them to be efficiently flushed to disk.

\SetKwFunction{KwCompact}{Compact}
\SetKwFunction{KwSortMapDV}{SortMapDV}
\SetKwFunction{KwDivide}{Divide}
\SetKwFunction{KwBuildTable}{BuildTable}
\SetKwFunction{KwUpdateOPD}{UpdateOPD}
\SetKwFunction{Kwupdateidtable}{UpdateIdTable}

\begin{algorithm}[]
{\small
\DontPrintSemicolon
  \KwInput{key columns $K_i$, encoded value columns $EV_i$ and  OPD $O_{i<V_i\leftrightarrow EV_i>}$ of merging \textit{SCTs} $S_i, i\in1,2,...,n$.}
  
  \KwOutput{key columns $K'_j$, encoded value columns $EV'_j$ and OPD $O'_j$ of new \textit{SCTs} after merge $S'_j, j\in1,2,...,n'$.}
  
  \tcp{key-value pair assembling, merge-sorting, garbage collection, and annotating SCT ID $s_i$ to a merged sequence}
  \mbox{\textit{$MergedSeq \{ <k',ev,s_i>\} \gets$\KwCompact($K_i, EV_i\ in\ S_i,i\in1,2,...,n$})};

  \tcp{Divide the merged sequence into $n'$ sub-sequences by size}
  \textit {$\{SubSeq_j<k',ev,s_i>,j=1,2,...,n'\} \gets$\KwDivide{$MergedSeq$}}\;
  
  \For{$SubSeq_j, j = 1,2,...,n'$}
  {
    \tcp{Construct a reverse index from the distinct value set 
    $V_j$ to the power set of <$EV\times S_i$> and sort by original values}
    
    \textit{$STReIndex_{\{V_j \to {2^{<EV\times S_i>}\}}} \gets$\KwSortMapDV{ $<k',ev,s_i>$ $ in\ SubSeq_j,$ $ O_{i\{V_i\leftrightarrow EV_i\}}$}}\;
    
    \mbox{\tcp{Constructing the new OPD based on the sorted reverse index}}
    $O'_{j\{V_j\leftrightarrow EV_j'\}}\gets $\KwUpdateOPD{$STReIndex$}
    
    \tcp{Build an index table from  $<EV\times S_i>$ to $EV'$}
    \textit {$EVtable_{\{<EV\times S_i>\to EV'\}} \gets $\KwBuildTable{$STReIndex, O_j'$}}\;

    \For{$entry<k',ev,s_i>$ in $SubSeq_j$}
    {
        $K_j'\gets K_j' \cup k'$;
        $EV_j'\gets EV_j' \cup EVtable[ev,s_i]$\;
    }
  \Return $K_j', EV_j',O_j'$ 
}
}  
\caption{OPD-based \textit{leveling} Compaction}
\label{alg:algorithm1}
\end{algorithm}

\begin{figure}[h]
\centering
\includegraphics[width=1\linewidth]{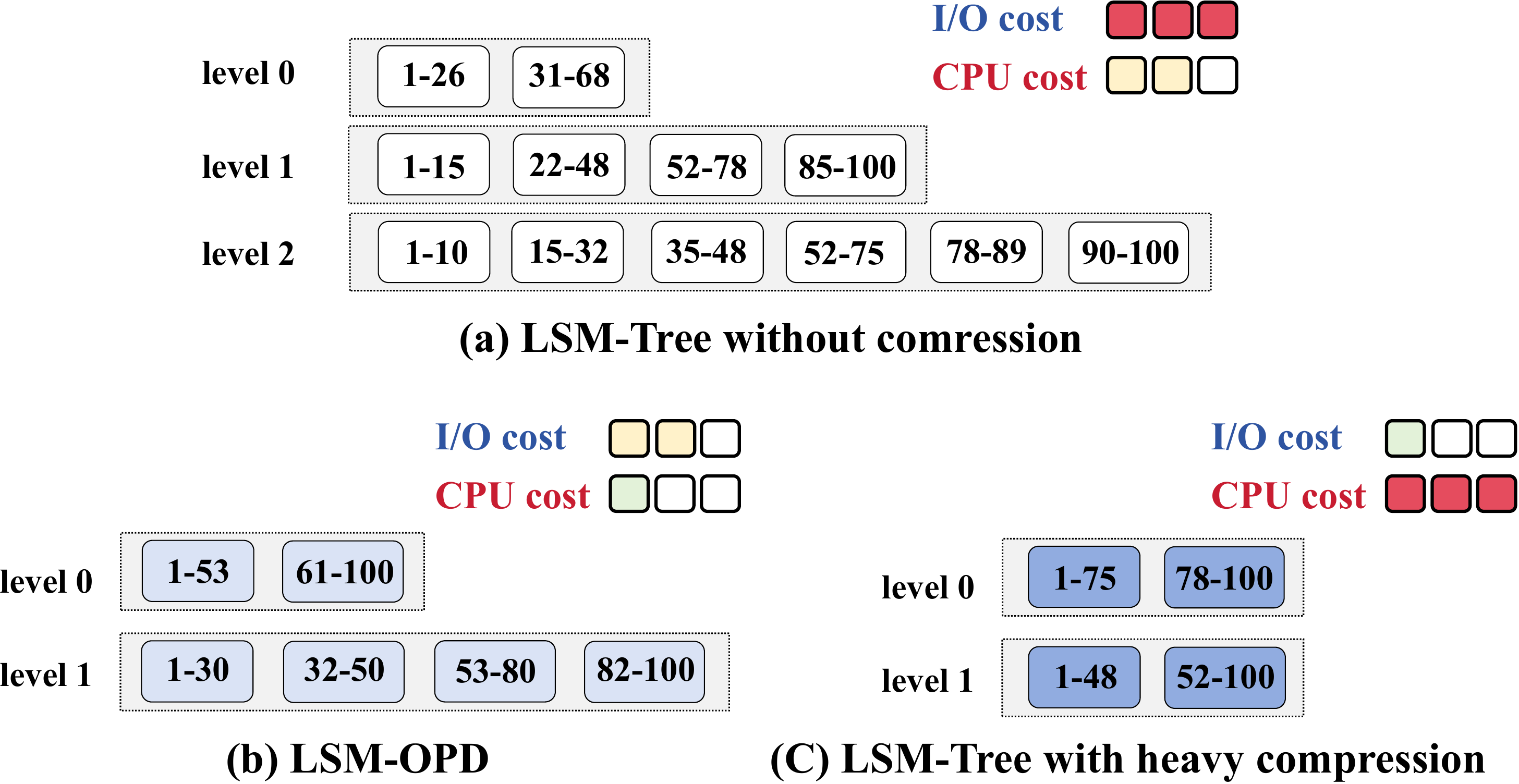}
\caption{Cost comparison of scan-based operations for LSM-Trees with different compression schemes}
\label{CC}
\end{figure}

\noindent \textbf{Cost Analysis of OPD-based Compaction.}
We proceed to conduct a detailed cost analysis of \textit{OPD-based} compaction in this section, comparing it to an LSM-Tree without any compression and one with heavy compression (\textit{e.g.,} LZ4, Snappy and GZIP compression algorithm), based on the term definitions provided in Table~\ref{term}.
Suppose that the same number of key-value pairs within an identical key range are uniformly inserted into an LSM-Tree with a fixed disk file size $F$. Due to the different degree of compression, the number of files and the total levels of the LSM-Tree may also be different, as shown in Figure~\ref{CC}. A higher degree of compression results in a more compact LSM-Tree with reduced height and depth, which in turn lessens the total number of file retrievals and I/O costs during compaction processes.
Specifically, for an LSM-Tree, the total I/O amount during compactions can be formalized as:
$C_{IO}=\sum_{i=1}^m F\cdot l_i\cdot T$, where $m$ refers to the total number of files.
And the $T$-times amplifications occur because each entry are averagely copied $T$ times at each level to be merged with the entries from the previous level, until it is propagated to the deeper level by the compaction process.
Thus, it's clear that the LSM-Tree with heavy compression scheme achieves the least total I/O costs during compactions at most cases due to the least number of disk files. And LSM-OPD follows closely, and potentially performing better when the NDV of inserted values is relatively low (\S\ref{EV}).

However, only reducing I/O cost by leveraging compression may produce very little effect, and could even have a negative impact on fast storage devises whose performance is mainly bounded by the CPU cost during compactions, since more recoding computation is traded for less I/Os.
Specifically, for an LSM-Tree without compression, the total CPU cost during compactions can be formalized~as:
\begin{equation*}
\begin{aligned}
&C_{CPU}=\sum_{i=1}^m (\frac{N}{m}\cdot S_k\cdot C_K + F\cdot C_C)\cdot l_i\cdot T 
\end{aligned}
\end{equation*}
Where the main costs are incurred by the merge sort of keys and the data copy in memory, which is also amplified by the factor of $T$.  While for an LSM-Tree with heavy compression, the total CPU cost is increased to:
\begin{equation*}
\begin{aligned}
&C'_{CPU}=\sum_{i=1}^{m'} [\frac{N}{m'}\cdot S_k\cdot C_K +F\cdot (C_C +C_D+C_E)]\cdot l_i\cdot T 
\end{aligned}
\end{equation*}
In cases where heavy compression is commonly applied to the entire file without considering data semantics, the cost of recompression is consequently incurred for the data size of the entire file. Although the number of files $m'$ under heavy compression is less than the one without any compression, the heavy recompression cost significantly dominates and further raises the overall CPU cost, as it is typically hundreds of times greater than the overheads of normal data copying.

In LSM-OPD, we alleviate this problem by offloading the heavy computation to lightweight OPDs and keeping large values encoded throughout the compaction process. As a result, LSM-OPD achieves a CPU cost of:
\begin{equation*}
\begin{aligned}
C''_{CPU}=\!\sum_{i=1}^{m''} (\frac{N}{m''}\!\cdot \!S_k\cdot C_K +F\!\cdot\! C_C+S_V \!\cdot C_S\!\cdot \!D_ilogD_i)\cdot l_i\cdot T
\end{aligned}
\end{equation*}
To simplify the analysis, the term $C_K$, $C_C$ and $C_S$ can be approximatively considered as small constants of the same magnitude and can be reduced during the fraction simplification, marked as $C$. Thus, we get:
\begin{equation*}
\begin{aligned}
C''_{CPU}&=\!A\!\cdot\! C\!\cdot\!\!\!\sum_{i=1}^{m''} (\frac{N}{m''}\!\cdot \!S_k +F)\!\cdot\! l_i\!\cdot\! T
,where\ A=\frac{S_V\!\cdot\! D_ilogD_i}{F\!\cdot\! (1+\frac{S_K}{S_K+S_O})}+1
\end{aligned}
\end{equation*}
With $F \approx \frac{N}{m''}\cdot (S_K+S_O)\approx \frac{N}{m}\cdot (S_K+S_V) $, it leads to:
\begin{equation*}
\begin{aligned}
\frac{C_{CPU}}{C''_{CPU}}&=\frac{\frac{N}{m}(2S_K+S_V)\sum_{i=1}^m\cdot l_i\cdot T}{A\cdot \frac{N}{m''}(2S_K+S_O)\sum_{i=1}^{m''}\cdot l_i\cdot T}\\
&=\frac{2S_K+S_V}{A\cdot (2S_k+S_O)}\cdot\frac{\sum_{i=1}^{d\cdot m{''}}l_i\cdot T}{d\cdot \sum_{i=1}^{m''} l_i\cdot T}
\end{aligned}
\end{equation*}
Where $d=\frac{m}{m{''}} \approx \frac{S_K+S_V}{S_K+S_O} $ refers to the total file compression ratio of LSM-OPD. Therefore, when the subsequent condition is satisfied:
\begin{equation*}
\begin{aligned}
\frac{2S_K+S_V}{A\cdot (2S_k+S_O)}\!>\!1\Leftrightarrow  \underbrace{D_ilogD_i<\frac{F}{S_V}\!\cdot\! \frac{S_V-S_O}{S_K+S_O}}_{I_1} 
\end{aligned}
\end{equation*}
we could obtain the follow results:
\begin{equation*}
\begin{aligned}
\frac{C_{CPU}}{C''_{CPU}}>\frac{\sum_{i=1}^{d\cdot m{''}}l_i\cdot T}{d\cdot \sum_{i=1}^{m''} l_i\cdot T} = \frac{\sum_{i=1}^{d\cdot m{''}}(\lceil log_T(i(T-1)+1)\rceil)\cdot T}{d\cdot\sum_{i=1}^{m{''}}(\lceil log_T(i(T-1)+1)\rceil)\cdot T}>1
\end{aligned}
\end{equation*}
The second part of
the inequality holds primarily due to the monotonic increase of the logarithmic terms in the numerator compared to those in the denominator. And in most practical cases that fit for dictionary encoding, the inequation $I_1$ holds. 
For example, consider a 32MB file that roughly accommodates up to $1,600,000$ OPD-encoded key-value pairs sized in $20$ bytes, $D_i$ must pass about $90,000$ to exceed the border of inequation $I_1$. To generalize, when the NDV ratio per file $\frac{D_i}{F/(S_K+S_O)}$ is below 5\%, which is much higher than the common NDV in the real datasets (e.g., 1\%~\cite{zeng2023empirical}), LSM-OPD  achieves more efficient compactions with a reduced complexity. And the border remains relatively stable regardless of the value size and file size, as shown in $I_1$.
The detailed analysis above has proven the superior performance of LSM-OPD in universally boosting essential scan-based compaction operations.
As shown in Figure~\ref{CC}, LSM-OPD not only significantly breaks the I/O bound by substantially reducing the total number of disk files, but also effectively optimizes the CPU cost by offloading the expensive computation of compactions to lightweight OPDs. As a result, LSM-OPD can achieve up to orders of magnitude improvement compared to an LSM-Tree without any compression or the one with heavy compression when processing scan-based compactions on values with relatively low NDV.

\subsubsection{\textbf{SIMD-based Vectorized Filter Evaluation.}}
When serving modern data-intensive applications involving HTAP, another crucial type of scan-based operations within LSM-Trees is the filter evaluation on values: $filtering(Value_{\{conditions\}})$.
Specifically, the essential process of filter evaluation is fundamentally similar to the compaction operation, which primarily involves merging sorted keys and identifying the correct versions of KV pairs. 
Thus, it exhibits similar performance bottleneck trends as compaction operations.
However, it encompasses a broader range of file retrievals across all levels and more complicated computations during filter processing, thereby presenting greater challenges for optimization. 

\begin{figure}[h]
\centering
\includegraphics[width=1\linewidth]{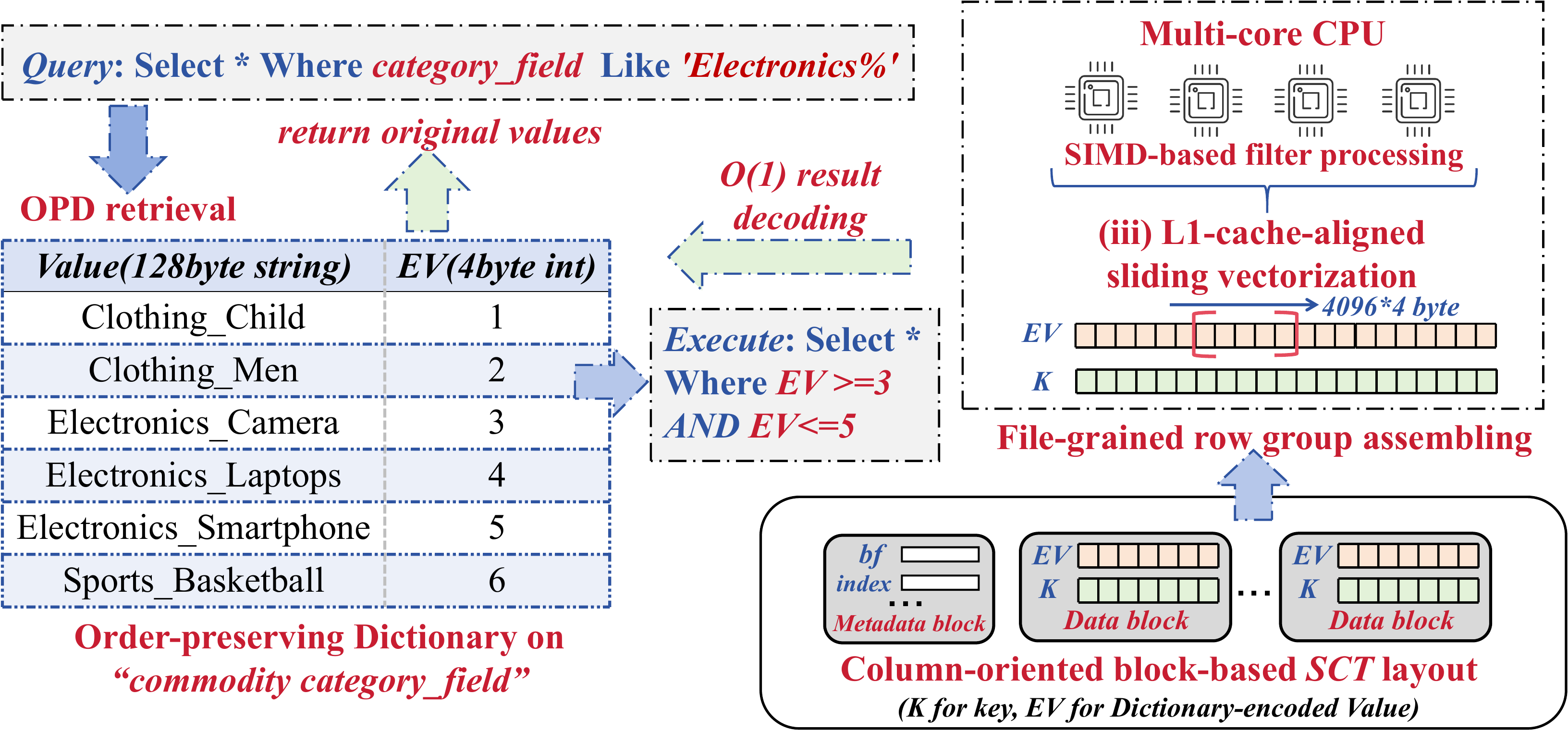}
\caption{Example of OPD-based vectorized filer evaluation}
\label{FE}
\end{figure}

LSM-OPD addresses the core issues by enabling direct filter evaluation on encoded data, thereby fully leveraging the parallel capability of modern multi-core CPU. Figure~\ref{FE} shows an example of OPD-based vectorized evaluation of a prefix query on $128\ byte$ string value of  ``\textit{commodity category\_field}''. First, the input prefix query on large string values will be transformed into a simplified range query on compressed numerical value by retrieving the order-preserving dictionary at a cost of $O(logD_i)$. Then, the filter is processed separately on each level of the LSM-Tree, and the results from each level will then be merged to discard the stale versions. Specially, for data in each \textit{SCT}, the block-wise $EV$ columns and $Key$ columns are assembled into a long sequence in memory without decoding. Where an $L1$-cache-aligned  vector sized in $4096\times4\ byte$ (total in $16 kb$ that fits in the $L1$ cache of CPU)
will slide over the sequence to copy data for further processing in CPU. In this way, the CPU is fully capable of parallelism  by leveraging the \textit{SIMD} instructions to evaluate filters directly on the encoded data. 
Then, the filtered results will be decoded back to the original values with $O(1)$ complexity, as the encoded values also refers to the offset of  original values in the OPDs. Finally, the query results from each level will be merged together to discard the stale versions and generate the correct outputs.

\noindent \textbf{Cost Analysis of OPD-based Vectorized Filter Evaluation.} Similarly, we conduct a detailed cost analysis of OPD-based vectorized filter evaluation in this section to demonstrate its high efficiency. As mentioned above, with the same number of inserted key-value pairs, the total number of disk files varies according to the different degrees of compression. As a result, the total I/O amount of evaluating a filter can be formalized as $C_{IO}=m\cdot F$, where $m$ refers to the total number of files. As shown in Figure~\ref{CC}, the result is consistent with the result of the cost analysis of compaction operation, where LSM-OPD performs the second best between the LSM-Tree without any compression and the one with heavy compression.

As for the CPU cost, an LSM-Tree without any compression evaluates a filter at a cost of:
\begin{equation*}
\begin{aligned}
C_{CPU}=N\cdot S_V\cdot C_S+r\cdot N\cdot [S_K\cdot C_K+ (S_K+S_V)\cdot C_C]
\end{aligned}
\end{equation*}
Where the first term refers to the string comparison cost of all inserted values, and the second term refers to the merge and the copy cost of qualified KV pairs, thus multiplied the selectivity of the filter. 
While the total CPU cost of processing a filter for an LSM-Tree with heavy compression is amplified to:
\begin{equation*}
\begin{aligned}
C'_{CPU}\!=\!m'\!\cdot\! F\!\cdot\! C_D+N\!\cdot\!S_V\!\cdot\! C_S+r\!\cdot\! N\!\cdot\! [S_K\!\cdot\! C_K\!+\! (S_K+S_V)\!\cdot\! C_C]
\end{aligned}
\end{equation*}
As it requires to decompress the entire file first in order to process the filter on original value semantics, resulting in expensive extra computations.

However, this cost is efficiently eliminated in LSM-OPD by enabling direct computing on encoded data, thereby further reducing the CPU cost to:
\begin{equation*}
\begin{aligned}
C''_{CPU}\!=\!\sum_{i=1}^{m{''}} logD_i\!\cdot\! S_V\!\cdot\! C_S+\!\frac{N\!\cdot\! S_O\!\cdot\!C_S}{S_I}\!+\!r\!\cdot\! N\!\cdot\! [S_K\!\cdot\! C_K\!+\! (S_K\!+\!S_V)\!\cdot\! C_C]
\end{aligned}
\end{equation*}
Where the first term refers to the decode cost of OPD retrieval, which operates at the $O(\sum_{i=1}^{m''} logD_i)$ level. And this cost is negligibly small compared to the $O(N)$ cost of  evaluating filters on large values.
While the second term refers to the SIMD-based vectorized filter processing, where a SIMD instruction can process $S_I$ size of data (\textit{e.g.,} 512 $bytes$) at a time. And the shared computational costs of merging and copying qualified KV pairs can be neglected in this comparison, since the selectivity of a filter $r$ could be extremely small (\textit{e.g.,} 1\%) in practice.

As a result, the capability of evaluating filters
directly on compressed data enables LSM-OPD to achieve an ideal performance improvement by a factor of $\frac{parallelism}{compress\_ratio}$, which is substantial. And the potential extra data movements caused by the misalignment between large string values and the CPU cache are also efficiently addressed by the sliding vectorization mechanism that directly operates on encoded data. In summary, LSM-OPD thoroughly boost the scan-based filtering operations by significantly breaking the I/O-bound and compute-bound at the same time for LSM-Trees on diverse storage devices. 

\section{EXPERIMENTAL EVALUATION}
\label{EV}

\subsection{Evaluation Setup}

\noindent\textbf{Competitors.} We comprehensively compare the performance of LSM-OPD with diverse LSM-based variants, including RocksDB$^1$
\footnotemark
v10.2.1~\cite{RocksDB}, RocksDB with heavy compression (snappy), RocksDB with key-value-separation  (BlobDB)~\cite{BlobDB,lu2017wisckey}, where values are separately managed in independent blob files, and BlobDB with Zstd-based dictionary compression operated on blob value files. 
The selected competitors encompass the mainstream LSM-based system designs that are capable of concurrent read/write processing with different degrees of compression.
\footnotetext{$^1$We disable the \textit{level\_compaction\_dynamic\_level\_bytes} optimization and the \textit{WAL} writes~\cite{RocksDB_Compaction} in RocksDB and set the $L_0$ file limitation for forced write stall to a consistent value across all systems, ensuring that all systems compact files in the same manner and that write stall is triggered under identical conditions.}

\noindent\textbf{Datasets and Workloads.}
To comprehensively evaluate the performance of LSM-OPD for a wide range of workloads that involves concurrent read/write processing and intensive scan-based operations, we extend the standard YCSB benchmark~\cite{cooper2010benchmarking} to support concurrent filter processing in LSM-Trees. The number of key-value pairs is set to $6.4\times 10^7$, whereas the size of entire dataset ranges from $2.86\ GB$ to $62.0\ GB$ according to different value sizes.
The sizes of keys and  files are set to $16B$ and $64MB$ respectively for all systems by default. Without special 
annotation, the NDV of value is set to 1\% by default, which covers more than 60\% real-world string values~\cite{zeng2023empirical}.
And the size, NDV and distribution of values, as well as
the selectivity of filters are varied in a reasonable range  according to the characteristics of real-world string dataset~\cite{zeng2023empirical} in subsequent detailed evaluations.

\noindent\textbf{Platform.} All experiments are performed on a 64-bit Ubuntu24.04 LTS workstation,  equipped with an Intel Xeon Silver 4314 CPU@ 2.40GHz and 512GB DDR4 RAM. For external storage, the workstation is mounted with 12TB HDD, 1TB SATA SSD, 4TB NVMe SSD, which can achieve up to about 180 MBs, 400MBps and 2300MBs sequential I/O performance. All codes are programmed in C++ and compiled using GCC v12.3.0 with O3 optimization. All systems are
allocated 16 threads by default, with 8 threads for front query processing and other
8 threads for background operations (e.g., compactions).

\begin{figure*}[h]
\centering
\includegraphics[width=1\linewidth]{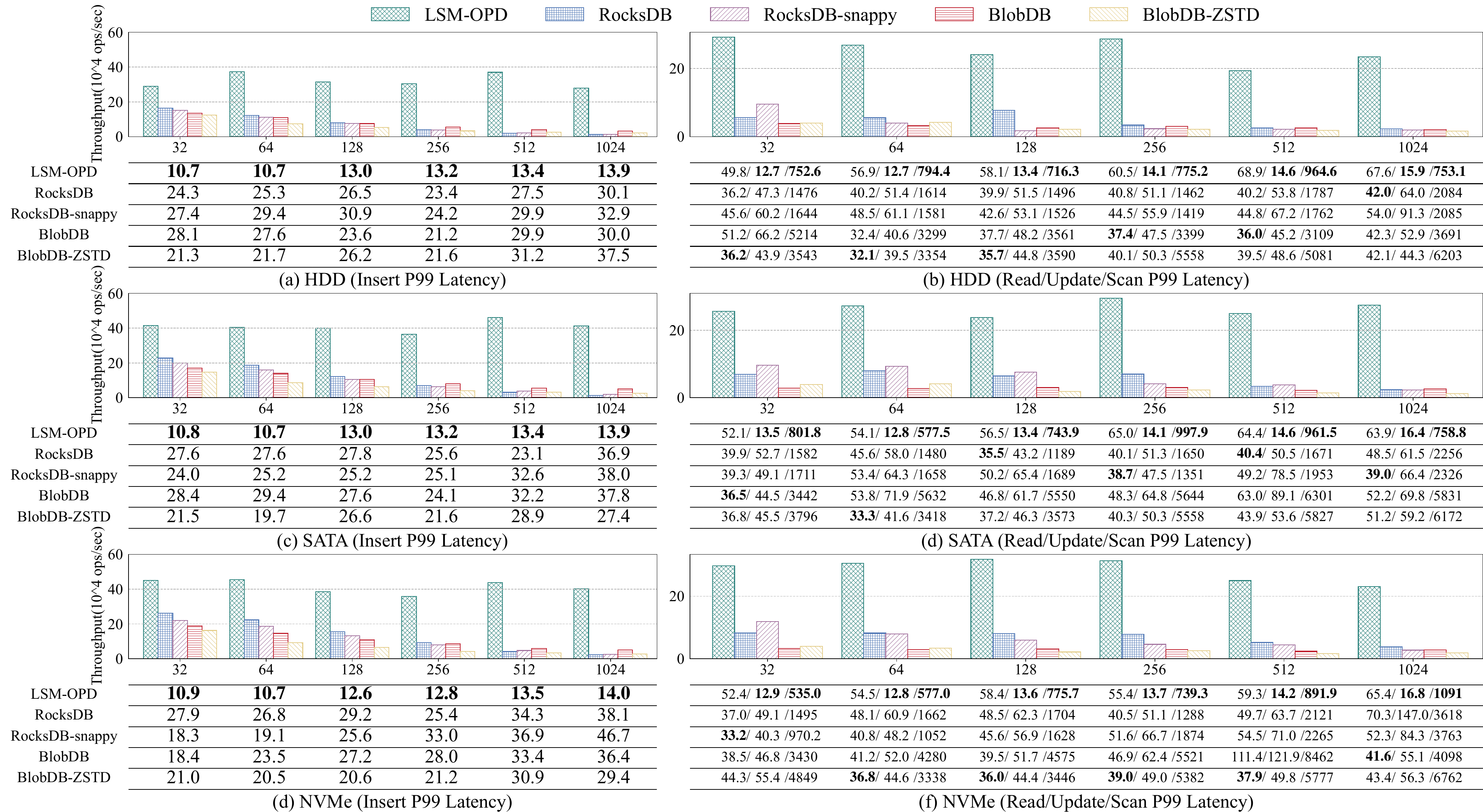}

\caption{The transactional processing performance comparison (P99 latency in microsecond)}
\label{TP}

\end{figure*}

\subsection{Transactional Processing Throughput}

\noindent \textbf{Key-Value Insertion.} We first evaluate the pure key-value insertion performance by varying 
value size from $32\ byte$ to $1024\ byte$, and fix the NDV and value distribution to $1\%$ and \textit{Uniform}, respectively.
As shown in Figure~\ref{TP}, LSM-OPD achieves averagely superior insertion throughput among all competitors especially for large value sizes. RocksDB optimizes the write amplification by employing the \textit{tiering} policy in $L_0$, and \textit{leveling} for deeper levels. However, when the number of tiered $L_0$ files reaches the limitation, a temporary stall will occur and lead to a throughput plunge. The write stall is more serious in the RocksDB with heavy compression due to the slower compactions, resulting in lower insertion throughput as shown in Figure~\ref{TP}. BlobDB achieves a better TP throughput when the value size is relatively large due to the key-value separation and pointer-based merge, which leads to lower write amplifications. However, despite faster decompression, similar performance degradation caused by compression is also reflected in BlobDB.
And BlobDB significantly scarifies the deletion persistence and the range lookup efficiency, let along filter processing.
Figure~\ref{TP} also shows that, as the value size increases, the TP throughput of other systems is correspondingly severely degraded due to the dramatically increase of I/O amounts and merge frequency. Although applying compression could alleviate the I/O pressure, the additional computational cost of decompression still significantly increases  as the value size grows, resulting in worse throughput and P99 latency.
Oppositely, despite the mild drawback caused by extra OPD constructing cost of  during flushes when the value size is small, LSM-OPD maintains excellent throughput across all value sizes. This is mainly attributed to the denser layout in \textit{SCT} files that always encodes large values as 4 byte integers.
And LSM-OPD further facilitates the compactions by enabling direct merge on encoded data without decompression, leading to minor I/O amounts, less compaction cost and frequency.  

\noindent \textbf{Hybrid Updates and Lookup Processing.}
Next we evaluate the transactional processing performance involving hybrid insertion, point read and short range lookup (averagely retrieve 500 adjacent key-value pairs) under the same value settings as above, where the ratio of updates, point read and range lookup is set to 50\%, 40\% and 10\%.
As shown in Figure~\ref{TP}, RocksDB exhibits a substantial decrease in overall throughput and increase in insert P99 latency in hybrid read/write test.
This is primarily due to 
the resources contentions (\textit{e.g.,} threads, bandwidth) between read/write processing and background compactions. And this issue becomes more pronounced as the value size increases, as evidenced by the rise in read P99 latency depicted in Figure~\ref{TP}.
While the mild OPD-based decode cost is emerged when the value size is relatively small (reflected in the P99 latency drawback of \textit{point\_read} in Figure~\ref{TP}),
LSM-OPD is mildly affected by the increase of value size and remains a superior overall throughput in all
transactional hybrid read/write processing. 
This is mainly attributed to the minor I/O and computations required by lookup and compaction operations in LSM-OPD, which largely benefit from the more compact \textit{SCT} file layout and the capability for direct computation on it.
Instead, RocksDB with heavy compression exhibits a degraded read performance as the value grows due to its heavy decompress costs. 
And BlobDB endures a more serious range lookup performance degradation due to the additional random value addressing in blob files. Additionally, more backstage resources are required by more complex garbage collection in blob files, which hinders the overall throughput as well. And heavy compression in blob files will further amplify these negative effects as the value size grows. As a result, LSM-OPD achieves higher overall throughput in transactional hybrid read/write processing, and exhibits lower average P99 latency in insertion, point lookups, and range lookups regardless of value sizes.

\begin{figure*}[h]
\centering
\includegraphics[width=1\linewidth]{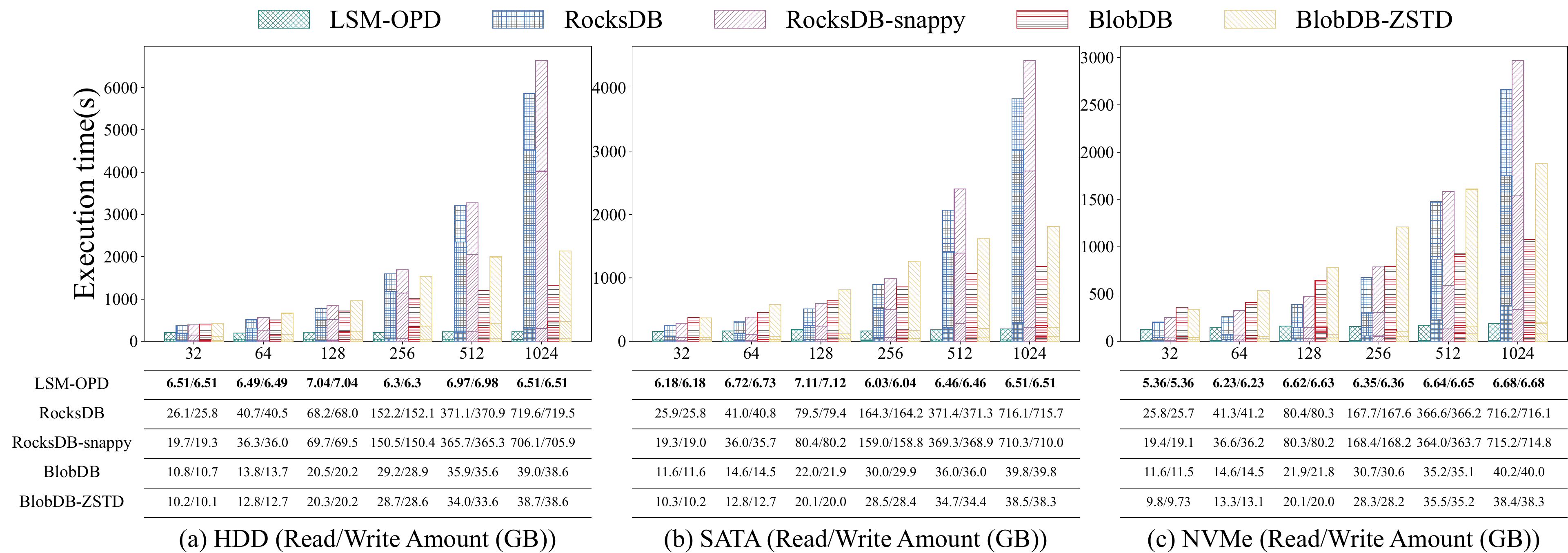}

\caption{The compaction performance comparison for different value sizes}
\label{Comp}

\end{figure*}

\subsection{Boosting Scan-based Operations}

\noindent\textbf{Compaction Performance.}
Compaction operations is source-intensive, and significantly affect the front query processing performance by determining the frequency and endurance of write stalls. In this section, we first 
evaluate the compaction performance of all competitors by varying the value sizes. 
As shown in Figure~\ref{Comp}, LSM-OPD costs least total time and I/O amounts on compactions under the same value configuration. This is mainly because the dense \textit{SCT} layout of LSM-OPD leads to less frequent compactions, and  enabling direct computing on compressed data facilitates more efficient compactions. BlobDB reduces the I/O costs by employing pointer-based value redirection in blob files during compactions to avoid expensive copies and sever write amplifications of large values, thus maintaining a relatively low and stable I/O amounts regardless of the compression degrees. While applying heavy compression on LSM-Trees could reduce the total I/O cost (which is even not pronounced in RocksDB, as the heavy compression is applied to the entire file that contain a mixture of key-value pairs and leads to a poor compression effect), it will increase the total time consumption due to the additional compute cost of decompression.

\begin{figure}[h]
\centering
\includegraphics[width=1\linewidth]{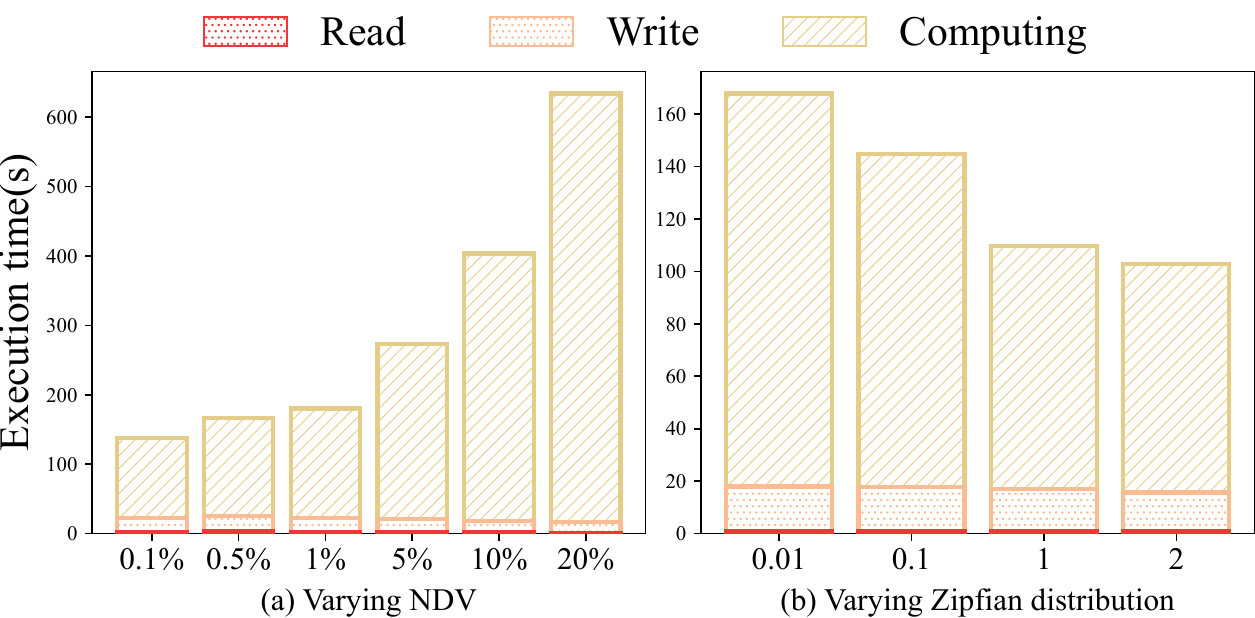}

\caption{The influence of different NDVs and distributions of values on the compaction performance of LSM-OPD (128 byte value)}
\label{NDV}

\end{figure}

We then
comprehensively evaluate the impact of different NDV and distribution settings of values on compaction performance during intensive insertions for LSM-OPD on representative SATA SSD, where the value size is fixed to 128 bytes.
As discussed before, the NDV and distribution of values may affect the compaction performance of LSM-OPD, since they collectively determine the NDV in each file $D_i$ and thus influence the size of OPDs. We first vary the NDV under a uniform value distribution. And then
the parameter $s$  of \textit{zipfian} distribution (fix NDV to 1\%) is varied to adjust the skewness of values: $p(k)=\frac{1}{k^s}/\sum^c_{n=1}\frac{1}{n^s}$, from 0.01 (uniform) to 2 (highly skewed), where $C$ denotes the total number of distinct values, and $k$ refers to the frequency rank. Better than expected,
despite a higher CPU cost when the NDV ratio in each \textit{SCT} file is roughly beyond 5\%,
LSM-OPD still 
maintains the superior overall compaction efficiency due to the significantly lower I/O amounts (Figure~\ref{Comp},\ref{NDV}).
And LSM-OPD retains this advantage until the NDV ratio roughly exceeds 20\% (compare to the results shown in Figure~\ref{Comp} (b) with 128 $byte$ value), which encompasses a range covering over 80\% of string columns in real-world datasets~\cite{zeng2023empirical}. 
And the total memory occupation of OPDs remains below 1 GB when the NDV is less than 10\%, indicating a relatively mild extra memory usage.
However, LSM-OPD may  experience performance degradation as the NDV increases and the value distribution becomes less skewed. In such cases, the general dictionary encoding method is also no longer recommended.

\noindent \textbf{Filter Performance.} 
In this section, we first evaluate the average performance of processing computationally expensive filtering operations on various value sizes ( selectivity is fixed to 1\% on uniform value distribution). Subsequently, we further analyze the influence of different NDV and value distributions, as well as filter selectivity, on filtering performance (fix value size to 128 bytes on SATA SSDs).
As a result, LSM-OPD significantly outperforms all competitors in all situations. RocksDB encounters a relatively high latency in filter processing due to the sever read amplifications in the multi-level structure and the expensive comparison cost of large string values. 
Applying heavy compression will cause  additional decompression cost and further increase the filter processing latency when the value size is small. However, the advantage of reduced I/O amount from a denser snappy-based compressed layout emerges as the value size grows in RocksDB.
As for BlobDB, it suffers from a worse filter processing performance due to the additional random value addressing in redundant blob files during filter processing. LSM-OPD addresses these issues by enabling vectorized filter processing  directly on compressed data. Where the large string values are compressed into a 4 bytes integers in an order-preserving manner, and SIMD-based parallelism is further employed to maximize the filter processing performance.
And the advantage of LSM-OPD becomes more pronounced when filter selectivity decreases, since the additional overhead, such as data copying cost, accounts for a smaller proportion of total cost, emerging the negative impact of inefficient filter processing.

\begin{figure*}[h]
\centering
\includegraphics[width=1\linewidth]{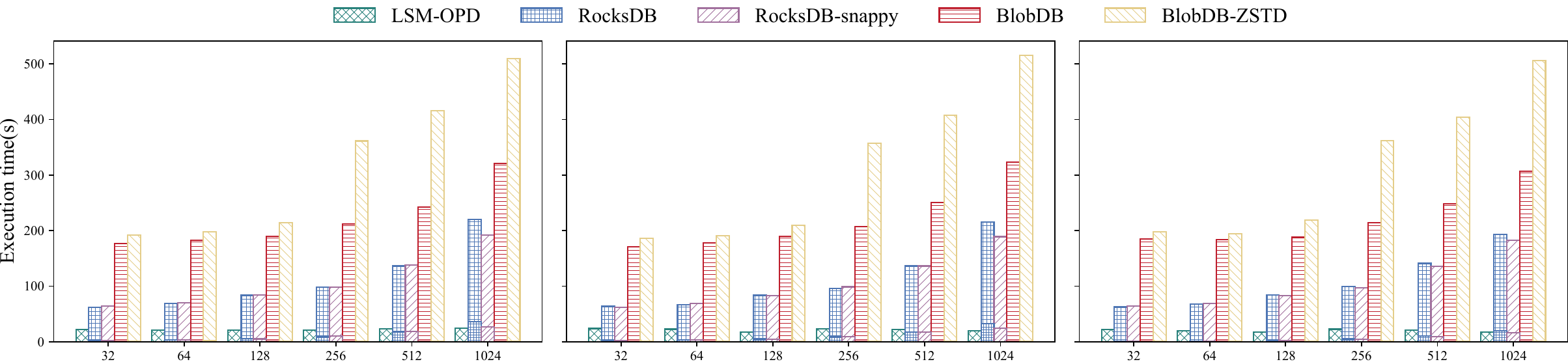}
\includegraphics[width=1\linewidth]{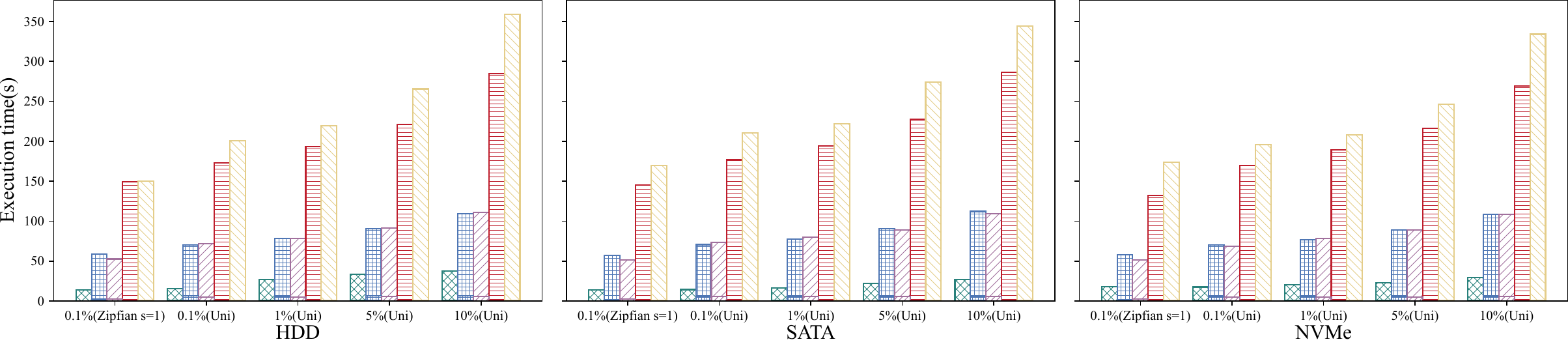}

\caption{The average filter processing performance varying value size and selectivity}
\label{AP}
\end{figure*}

\begin{figure*}[h]
\centering
\includegraphics[width=1\linewidth]{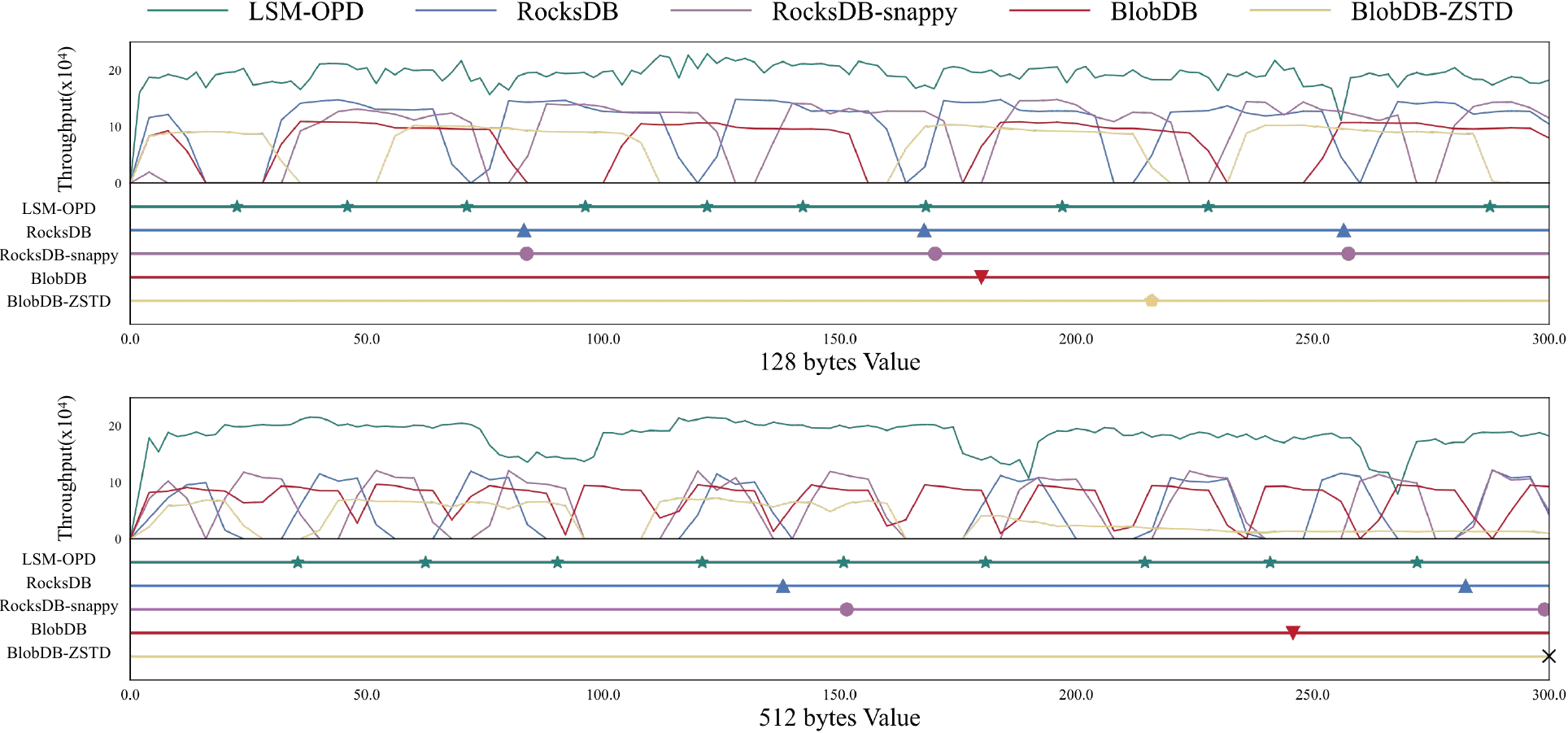}

\caption{The HTAP performance for different value sizes on SATA SSD}
\label{HTAP}

\end{figure*}

\subsection{Hybrid Transactional/Analytical Processing}

In this section, we verify LSM-OPD's advantage in intensive hybrid transactional and analytical processing. Due to space limitations, we select two representative value sizes and conduct experiments on SATA SSDs.
Each system is allocated 16 threads, with half dedicated to front-end query processing (including 4 threads for transactional processing, and the other 4 for filter processing) and the other half assigned to back-end operations (\textit{e.g.,} compactions). During the HTAP experiment, transactions are continuously executed after the insertion of $6.4\times 10^7$  key-value pairs, with a workload consisting of  50\% updates, 40\% point reads and 10\% range lookups,  concurrently
with intensive filter evaluations. This workload is highly challenging for LSM-based systems, as both front-end queries and background threads require computational resources to process intensive scan-based 
operations. Without specific optimization, it may lead to chaotic resource contentions (e.g., I/O bandwidth and CPU cores) and result in sever stalls. To illustrate the impact of hybrid workloads on the system more intuitively, we plot the curve of TP throughput as it varies with time in 300s, along with the latency of each filter on the timeline in Figure~\ref{HTAP}. As a result,
all competitors face different degree of TP throughput degradations and write stalls when concurrently processing filters. Among these, despite its excellent compaction performance, BlobDB exhibits the lowest TP throughput and AP latency due to its poor scan performance. RocksDB maintains a moderate TP throughput and AP latency, but experiences a dramatic throughput decrease when encountering sever write stalls.
These drawbacks are more pronounced when the value size getting larger, as both compaction and filtering operations demand more compute resources. While LSM-OPD achieves a comprehensive optimization of scan-based operations, including efficient backstage compaction and front filtering, by enabling direct computation on compressed data, thereby significantly reducing the frequency of write stalls and the latency of filtering.

\section{RELATED WORK}

\noindent \textbf{Offloading Computation within LSM-Trees to Heterogeneous Devices.} LSM-Trees heavily rely on frequent backstage compactions to perform physical data modifications, which is resource-intensive and may lead chaotic CPU resource contentions with front query processing under intensive concurrent read/write processing. A number of studies focus on alleviating this issue by offloading the heavy computation within LSM-Trees to other modern computing devices, such as GPUs~\cite{zhou2024gpu,sun2025rgkv,sun2024glsm,xu2020luda}, FPGAs~\cite{zhang2020fpga,sun2020fpga,im2020pink} and other specialized hardware like DPUs~\cite{ding2023dcomp,ding2024d2comp} and NDP structures ~\cite{sun2025prockstore}. 
Specifically, LUDA~\cite{xu2020luda} leverages GPUs to process SSTables via a co-ordering mechanism, which  minimizes data movement and consequently alleviates CPU pressure. gLSM~\cite{sun2024glsm}, on the other hand, decouples keys and values to reduce data transfer between the CPU and GPU, thus enhancing the efficiency of compactions. 
OurRocks~\cite{choi2020ourrocks} speeds up the  analytical processing within LSM-Trees by offloading the disk scan directly to GPUs.
Sun et al.~\cite{sun2020fpga} and Zhang et al.~\cite{zhang2020fpga} propose an accelerated solution for key-value stores by offloading the compaction task to an FPGA, thereby reducing CPU resource contention in a more economical way. 
dCom and d$^2$Com~\cite{ding2023dcomp,ding2024d2comp} reduce CPU overload by leveraging DPUs to accelerate the compaction of SSTables based on a heterogeneous structure.
However, heterogeneous computing still necessitates extra data transfer from main memory to other computing units, which may potentially limit overall system performance due to increased latency and bandwidth constraints. And relying on specialized computing devices may significantly increase the cost of deploying and operating data service clusters  while also leading to inefficient utilization of computing resources.
LSM-OPD aims at achieving the same goal in a more cost-effective and ingenious manner by enabling direct computation on compressed data within LSM-Trees,  eliminating the need for additional specialized computing devices.

\noindent\textbf{Optimizing Scan-based Operations within LSM-Trees.} Another trend of research focus on optimizing the scan-based operations within the original design space of LSM-Trees to reduce write amplifications and enhance read performance.  Dayan et al.~\cite{dayan2018dostoevsky,dayan2019log}
collaborates the \textit{tiering} and \textit{leveling} policy at level granularity to
achieve better performance trade-offs. Sarkar
et al.~\cite{sarkar2021constructing} systematically construct and analyze the design space of compaction
policy for LSM-based systems. Huynh et al.~\cite{huynh2022endure,huynh2024towards} focus on tuning the parameter configurations within LSM-Trees under workload uncertainty.
While Luo et al.~\cite{yu2024camal,mo2023learning} introduce machine learning-based parameter tuning paradigm for the optimal layout in LSM-Trees.
As for scan performance within LSM-Trees, real-time LSM-Tree~\cite{saxena2023real}
firstly utilizes the compaction process to change the storage layout
from row-based to column-oriented in order to accelerate relational HTAP for multi-column data. Disco~\cite{zhong2025disco} introduces a compact secondary index structure to boost range lookup operations by pruning unnecessary file retrievals in LSM-Trees. LSM-OPD presents a brand new solution to thoroughly boost scan-based operations by enabling direct computing on compressed data, simultaneously breaking the I/O bound and CPU bound. As a result, LSM-OPD is essentially orthogonal yet
collaborative with the above techniques.  

\noindent \textbf{Compression and Vectorization Techniques.} Compression techniques have been widely studied in the database community, especially for column-oriented storage. Abadi et al.~\cite{abadi2006integrating,abadi2008column} has laid the foundation of columnar-oriented compression scheme for databases, and introduced the trade-offs between different compression techniques. Zeng et al.~\cite{zeng2023empirical} empirically evaluate various column-oriented compression schemes in modern applications and verify the high efficiency of dictionary decoding in real-world large string datasets. Liu et al.~\cite{liu2019mostly} introduce a probabilistic mostly order dictionary encoding scheme to enhance the range filter evaluation directly on encoded data when facing uncertain data modifications. As for vectorized execution, Kersten et al.~\cite{kersten2018everything} systematically introduce the vectorized execution scheme and discuss its efficiency on various value sizes. VIP~\cite{polychroniou2020vip} built a vectorized analytical query engine and emphasize the importance of cache alignment. However, most of the column-oriented compression schemes fail to support efficient recoding in the face of intensive data updates, and may resulting in severe temporary stalls. LSM-OPD not only aims at leveraging the power of direct vectorized computing on compressed data to boost scan-based operations in LSM-Trees, but also introduces a novel LSM-based framework  of dynamic lightweight data encoding for HTAP workloads.

\section{CONCLUSION}
This paper presents LSM-OPD, a Log-Structured Merge-Order-Preserving Dictionary encoding scheme that enables direct computing on compressed data within LSM-Trees. At the core of the LSM-OPD is offloading the heavy I/Os and computation on large string values to lightweight OPDs and fully leverages the power of SIMD-based parallelism of modern multi-core CPUs. 
As a result,
it thoroughly boosts the scan-based operations by achieving more efficient compaction and filtering with a significantly reduced I/O amounts and optimized computational complexity, especially when the NDV is relatively low. Extensive experiments demonstrate the superior
efficiency of LSM-OPD in processing various workloads  that involve intensive scan-based operations on diverse modern storage devices.

\section{acknowledgements} This work was partially supported by CCF-Huawei Populus Euphratica Challenge Research Funding (No.CCF-HuaweiDB202406),   National Natural Science Foundation of China grant 62372138, Natural Science Foundation of Heilongjiang Province of China grant HSF20230095 and 2024ZXJ01A04.


\newpage

\balance
\bibliography{sample-base}

\end{document}